# SLEGO: A Collaborative Data Analytics System with Recommender for Diverse Users

Siu Lung Ng, Hirad Barada Rezaei and Fethi Rabhi

School of Computer Science and Engineering, The University of New South Wales, Sydney, NSW 2052, Australia
{siu_lung.ng, hirad.rezaei, f.rabhi}@unsw.edu.au

**Abstract.** This paper presents the SLEGO (Software-Lego) system, a collaborative analytics system that bridges the gap between experienced developers and novice end users using a cloud-based platform with modular, reusable microservices. These microservices enable developers to share their analytical tools and workflows, while a simple graphical user interface (GUI) allows novice users to build comprehensive analytics pipelines without programming skills. Supported by a knowledge base and a Large Language Model (LLM) powered recommendation system, SLEGO enhances the selection and integration of microservices, increasing the efficiency of analytics pipeline construction. Case studies in finance and machine learning illustrate how SLEGO promotes the sharing and assembly of modular microservices, significantly improving resource reusability and team collaboration. The results highlight SLEGO's role in expanding access to data analytics by integrating modular design, knowledge bases, and recommendation systems, thereby fostering a more inclusive and efficient analytical environment.

## 1 Introduction

In the era of big data, the ability to analyze both structured and unstructured data has become essential for organisations, researchers, and individuals aiming to improve decision-making, uncover patterns, and address complex problems [1]. However, a significant divide persists between users proficient in programming and those who are not. Traditional approaches rely on business users to outline requirements. Software engineers then translate these into specialized tools, creating inefficiencies and limiting flexibility [2].

While tools like Microsoft Excel provide accessible analytical capabilities, they often struggle with ensuring accuracy and reliability for more complicated analysis, especially for users with limited technical expertise [2, 3]. Professionals such as accountants often resist transitioning to more sophisticated analytics platforms, preferring familiar tools at the cost of missing out on the potential of newer technologies [3]. In contrast, advanced tools like Python and Jupyter Notebook empower technical users to develop complex, reproducible analytical workflows by integrating explanatory text, code, and outputs in a single environment [4]. However, these tools demand significant programming expertise, creating a steep barrier for non-technical users.

This dichotomy highlights a broader challenge: while powerful tools require advanced programming skills, simpler tools lack the flexibility and functionality needed for complex analytics [2, 5]. Bridging this divide—by creating a system that combines advanced analytical capabilities with accessibility for non-technical users—remains an ongoing challenge, especially for general-purpose applications designed to support diverse use cases.

To bridge the divide, we introduce SLEGO (Software-Lego), a collaborative analytics system that integrates an intuitive graphical interface with advanced computational capabilities. The SLEGO architecture establishes a unified framework where technical experts can develop modular microservices, domain specialists can contribute knowledge, and novice users can construct sophisticated analytics pipelines without programming expertise. Through a comprehensive low-code platform enhanced by a knowledge base and LLM recommendations, SLEGO broadens access to advanced data analytics while promoting code reusability and cross-team collaboration. This paper examines SLEGO's design principles and implementation, demonstrates its effectiveness through case studies, and discusses future directions. The system is available as an open-source project to foster continued innovation in collaborative analytics. The paper introduced the problem and need for bridging skill gaps (Section 1), surveyed related literature and identified missing capabilities (Section 2), presented SLEGO's architecture and implementation (Sections 3 and 4), demonstrated its effectiveness via case studies (Section 5), discussed improvements (Section 6), and concluded with the system's significance and future directions (Section 7).

## 2      Background and Literature Review

Collaborative Data Analytics (CDA) is essential in fields ranging from emergency management to IoT data analysis, yet it faces significant technical challenges. Tucker et al. (2017) identify data management as one of several technical challenges in collaborative data analytics for emergency response, including issues of data access and privacy. The paper also highlights information sharing capabilities between different systems as a technical consideration, while noting that no single computational model has emerged as an industry standard. These technical factors are presented as part of a broader framework that includes skills-related, organizational, and political challenges. These factors highlight the urgent need for sophisticated frameworks to manage the complexities of CDA applications efficiently [6]. Banerjee and Chandra (2019) developed a platform to enhance Collaborative Data Analytics (CDA) across IoT domains, featuring a language-agnostic design for simplicity, addressing issues like tool heterogeneity and version management, akin to DataHub's single-platform approach by Bhardwaj et al. (2015) [7, 8]. Callinan et al. (2018) investigated the role of collaborative platforms in co-creating value from open data, highlighting essential factors like participant characteristics and trust [9]. Yamakami (2020) noted the need for research into systematic management of analytics handovers due to their complexity and the unique needs of individual analysts [10].

In organisational settings, data analytics faces critical challenges centered on component utilisation and flexibility across different expertise levels. While developers create reusable analytical functions for multiple applications, these often remain underutilized, leading to redundant development efforts [11]. This inefficiency particularly impacts novice and business users, who struggle with inflexible tools that resist modification despite being technically accessible [2]. The sharing of analytics resources across teams presents additional complexities [12, 13], with disparate software and hardware configurations limiting function transferability between domains like finance and retail forecasting. This situation necessitates a centralized analytics platform that can serve users ranging from those using basic web applications to advanced programmers requiring sophisticated customisation capabilities [12]. The challenge extends to identifying and effectively reusing existing code and abstractions [14]. Low-Code Platforms (LCPs) have emerged as a promising solution for streamlining software development. As Bock and Frank (2021) note, these platforms effectively integrate traditional tools into user-friendly environments, though questions about their maturity persist. Their research emphasizes the need to evaluate LCPs' potential in improving conceptual modeling practices throughout the software development lifecycle [15].

Interactive notebook-based tools—such as Jupyter Notebook [16], DeepNote [17], and Google Colab [18], —have gained popularity for data analytics but generally require programming expertise. DASE architecture [19] proposes a knowledge base approach to support data analytics, while systems like Alteryx [20], KNIME [21] and RapidMiner [22] are low-code platforms for data analytics but missing knowledge base farmwork to support analytics. Liu et al. (2022) pointed out that Texera supports collaborative editing and scalable computation, reflecting the evolving needs in collaborative data analytics tools [23, 24]. Building on these developments, Ng et al. [25] demonstrated an architecture integrating data acquisition, analytics microservices [26], and cloud-based visualisation. Their implementation of the ADAGE framework [27] showed how organisational design guidelines can facilitate modular, service-oriented component development, making complex analyses accessible to finance domain experts without requiring advanced technical skills.

The integration of LLM and Artificial Intelligence (AI) is rapidly transforming the data analytics and software development landscape [28, 29]. In natural language processing, tools like Langchain [30] provide modular architectures for seamless LLM integration, while RAG (Retrieval-Augmented Generation) [31] enhances performance through external knowledge retrieval. These technologies demonstrate significant potential for creating sophisticated, context-aware AI systems. Recent research has showcased diverse applications of LLMs. Dai, Xiong, and Ku (2023) presented a framework leveraging LLMs to enhance qualitative research efficiency, potentially surpassing traditional human coding capabilities [32]. Liu et al. (2023) developed JarviX, a platform using fine-tuned LLMs to guide users through data visualisation and machine learning optimisation, making advanced analytics accessible to non-technical users [28], however, the platform does not focus on collaboration for diverse types of users. Additionally, Nejjar et al. (2023) explored LLMs' application in code generation and data analysis, highlighting both their potential for enhancing research productivity and the need to address output integrity concerns [33]. Furthermore, some popular data analytics tools, such as MS Excel, have embedded LLM or AI features to enhance their analytical capabilities [29]. These developments collectively point toward a future where AI significantly enhances information retrieval, content generation, and decision-making processes across multiple domains, while maintaining awareness of the need for careful validation and quality control.

As a conclusion of literature review, there is a lack of integrated approaches that combines the advantages of low-code platforms, knowledge bases and LLM/AI methods as well as enable collaboration between diverse types of users to facilitate the development of complex data analytics solutions. Such an approach would enable modularity, reusability and data analytics knowledge sharing between novice and expert users, reducing the barriers for cross-disciplinary collaboration.

## 3     The Design of SLEGO System

### 3.1    Basic Principles

The proposed SLEGO architecture addresses the challenges outlined earlier by integrating a modular microservices design, shared knowledge base and LLM/AI-driven recommendations into a low-code platform. This approach simplifies analytics pipeline construction, enhances collaboration, and bridges expertise gaps, making advanced data analytics more intuitive and accessible to a broad user base. The platform bridges the skill gap between seasoned experts and beginners by identifying three important roles: **1. Technical Experts:** create and maintain microservices that form the system's computational foundation. They develop modular code following standardized guidelines, validate functionality, and ensure system compatibility. Their role enables the continuous expansion of the platform's analytical capabilities through reusable components. **2. Domain Experts:** enhance the system's analytical capabilities by reviewing microservices and enriching them with domain-specific knowledge. They create example pipelines that demonstrate effective analytical workflows, which are stored in the knowledge base as templates. This contribution is essential for guiding future users and improving the recommendation system's accuracy. **3. End Users:** interact with the system through two primary pathways: direct pipeline construction or recommender-assisted development. In the direct approach, users select and configure microservices from the component library. Alternatively, they can submit queries to the LLM recommender, which suggests pipeline configurations based on knowledge base matches. Together, these collaborative features achieve three key objectives: (1) providing novice end users with access to advanced analytics, (2) facilitating knowledge sharing and reuse of analytics microservices, and (3) enabling AI-assisted analytics for users with limited technical expertise.

### 3.2   SLEGO Architecture

The SLEGO system is supported by a modular architecture that prioritizes reusability and customisation while maintaining system cohesion. **Fig. 1** illustrates the system's core components and their interactions.

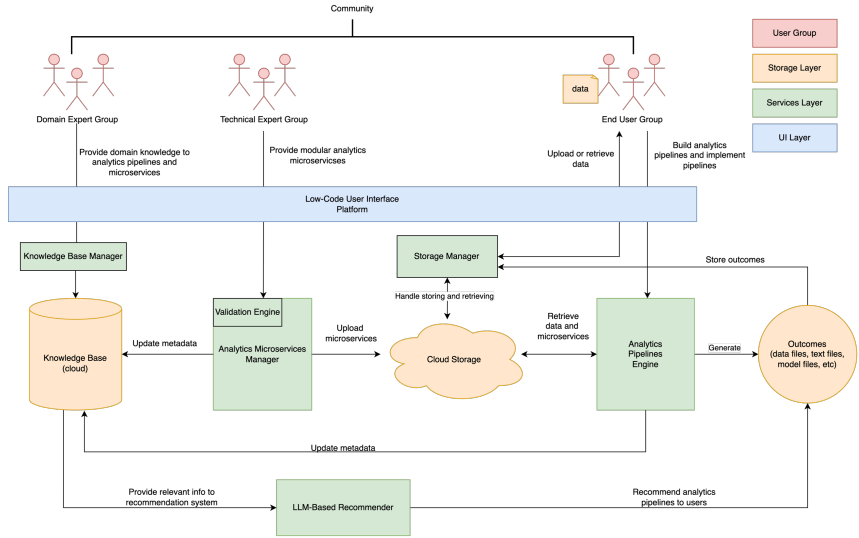

**Fig. 1.** SLEGO Architecture

**Storage Layer**
**Cloud Storage**: Ensures secure and scalable access to all analytics components and data. It supports collaboration, resource sharing, and seamless system integration, enabling unified resource management and uninterrupted system operation.
**Knowledge Base**: Serves as a centralized metadata repository housing references to both analytics microservices and user-constructed pipelines. Domain experts enrich these components by embedding essential domain knowledge, including detailed documentation, practical use cases, and technical limitations, all written in accessible language for end users and LLM interpretation. This repository not only preserves model elements but also functions as an educational resource through its collection of existing pipelines. By providing structured metadata and domain expertise, it enables the recommendation system to optimize analytics processes effectively. Below are the key components of the knowledge base:

- **Microservices Repository**: **(**a**) Microservice Name**: Unique identifier for each analytics microservice defined by developers. (b) **Description**: Provides a detailed description of what the analytics microservice does, defined by domain experts. (c) **Docstring/Code Comment**: Contains detailed documentation embedded within the microservice. This includes usage examples, parameter explanations, and other notes that help developers and users understand the specific operations performed by the microservice. **(d) Parameters**: Lists parameters names and values **(e) Source Code**: Contains microservice code. **(f) linked Pipeline**: the pipeline(s) uses these services.
- **Pipelines Repository: (a) Pipeline Name**: Defined by users, through the GUI. **(b) Description**: Elaborated by domain experts, which may include human experts or LLM-powered models. **(c) Microservices Details**: Details the sequence of microservices operations in the pipeline. **(d) Embedding**: This column stores vector representations of pipeline descriptions. These vectors enable the computation of similarities between stored pipelines and user queries for the recommendation system.

**Data and outcomes**: Users can upload data in various formats to SLEGO's Cloud Storage, and Analytics Pipelines Engine generates diverse outcomes/files, such as processed data, visualisations, logs, and models, which are stored in cloud storage for reuse and future workflows.

**Service Layer**
**Analytics Microservices Manager and Validation Engine:** The Analytics Microservices Manager handles the upload of modular and reusable microservices to the Cloud Storage. Before uploading microservices, the Validation Engine is triggered to ensure all microservices meet basic executable standardized, including structure, input parameters, documentation, and output. Only validated microservices are accepted for upload, maintaining platform integrity and ensuring seamless integration into analytics pipelines. Microservices are designed with the following principles:

- *Standardisation*: Modular design with standardized coding practices.
- *Flexibility*: Configurable parameters and input/output paths for flexible data access.
- *Output as a File*: Generates files/objects for integration with other microservices.
- *Data Accessibility*: Stores data in shared cloud space for easy access.
- *Documentation*: Includes clear docstrings or comments to clarify functionality.
- *Naming Convention*: Uses intuitive and human-readable names for compatibility with users and the LLM recommender.

Note that Analytics Microservices can also wrap external APIs or programs from other languages, extending their functionality while adhering to SLEGO's standards.
**Validation Engine**: Ensures all uploaded microservices meet standardized guidelines by validating structure, input parameters, documentation, and output. This maintains platform integrity and enables seamless integration into analytics pipelines.
**Storage Manager** and **Knowledge Base Manager:** These components manage the organisation and retrieval of data for Cloud Storage and Knowledge Base.
**Analytics Pipeline Engine:** Orchestrates the structured flow of data through multiple microservices, transforming raw data into actionable insights with consistency and reproducibility (see Section 3.4 for details).
**LLM-Based Recommender:** Leverages the knowledge base to provide pipeline-building suggestions based on user queries. This assists end users in constructing analytics pipelines efficiently (see Section 3.5 for detailed workflow).

**UI Layer**

**Low-Code UI**: SLEGO's GUI offers a straightforward interface that unites technical experts, domain experts, and end users. Domain experts can update the knowledge base with domain-specific insights, while technical experts manage and refine microservices. End users can build and run analytics pipelines by simply selecting components, adjusting parameters, and executing their workflows—no coding required. This user-friendly interface emphasizes insight generation over technical complexity. Its workflow involves three main steps: (1) Selecting microservices from a dropdown menu, (2) entering required parameters, and (3) executing the pipeline with a button click.

### 3.3 User Interaction and Workflow

SLEGO provides a number of use cases for the three user groups which involve specific system components. Technical experts primarily interact with the Analytics Microservices Manager and Validation Engine when uploading new microservices to Cloud Storage, ensuring standardized implementation. Domain experts utilize the Knowledge Base Manager to review microservices and pipelines, enrich them with domain expertise documentation, and create exemplar pipelines as templates for others.

The main use case for end users, as illustrated in the sequence diagram shown in **Fig. 2**, encompasses three phases. In the data management phase, users can optionally upload datasets through the interface, which are then processed by the Storage Manager and stored in Cloud Storage for analysis. During pipeline construction, users either directly select and configure analytics services from the component library or receive AI-assisted suggestions from the recommendation system based on their requirements. In the execution phase, the Analytics Pipeline Engine retrieves the necessary microservices from Cloud Storage, performs the specified computations, and stores the results back in Cloud Storage. The Storage Manager then retrieves these results, which may include new data files, visualisations, or analysis outputs, and forwards them through the interface for user review. Users can then save successful pipeline configurations as templates in the Knowledge Base for future use.

This integrated approach ensures systematic knowledge sharing while maintaining system quality and accessibility across different expertise levels, demonstrating how SLEGO combines technical expertise, domain knowledge, and user interaction in a cohesive analytics platform.

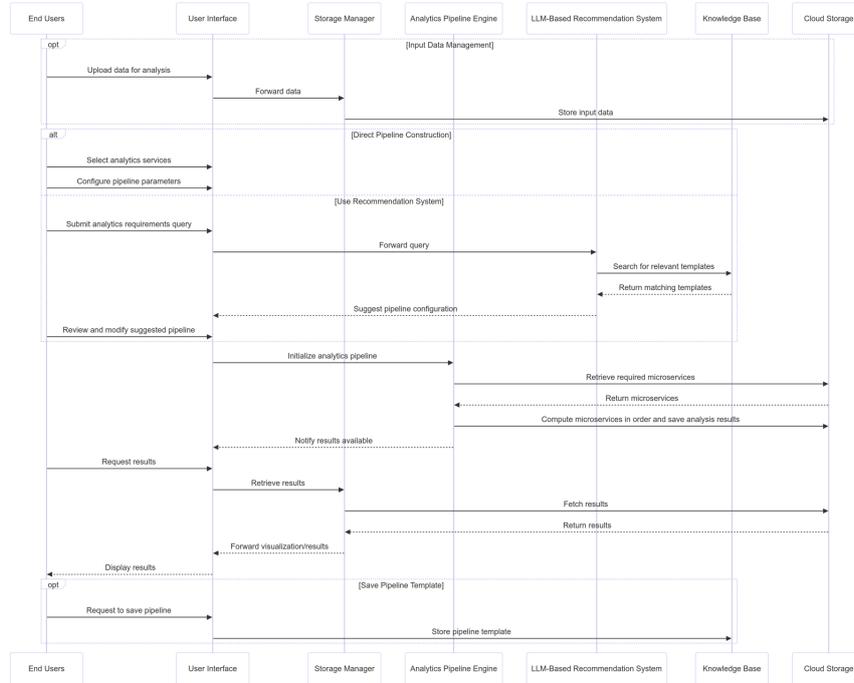

**Fig. 2.** Use Case "Data Analytics Pipeline Construction and Execution" Sequence Diagram

### 3.4   Analytics Pipeline Engine

The Data Analytics Pipelines Engine manages the structured flow of data through multiple microservices, transforming raw data into actionable insights with consistency and reproducibility. **Fig. 3** illustrates the process of how the engine transform the input data with multiple microservices to get the final output. These pipelines handle tasks ranging from data ingestion and cleansing to complex computations. Microservices operate sequentially as defined in the configuration, processing tasks and generating output files stored in the cloud. These outputs can serve as inputs for subsequent microservices, culminating in a final data file that meets the user's specific requirements.

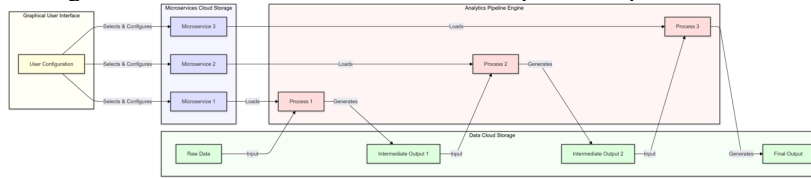

**Fig. 3.** SLEGO Microservices Workflow Diagram

### 3.5 LLM-Based Recommendation System

This system acts as a virtual assistant, fine-tuning the selection of analytics pipelines and parameters based on specific user requirements. The design was inspired by RAG [31] and multiagent-LLM approach [34]. Each step of the recommendation process is detailed in **Fig. 4** and follows a structured operational flow within the SLEGO system:

1. **Query Input and Incomplete Pipeline Input:** Users start by inputting a detailed analysis request or an incomplete pipeline. The system offers three interaction methods: (a)Submitting a query with analysis requirements. (b) Providing an incomplete pipeline for completion recommendations. (c) Combining both methods.
2. **Embedding:** The system processes the query embeds into a numerical vector.
3. **Matching:** The query vector is compared against pipeline descriptions in the knowledge base using cosine similarity, identifying the best match pipelines
4. **Selection of Top N Pipelines:** The system extracts the top N matching pipelines and their microservice details from the knowledge base based on similarity scores.
5. **AI Pipeline Recommendation Prompt:** The selected pipeline details, user query, and incomplete pipeline input are transformed into a structured text prompt for LLM input. This prompt includes rationale for each recommendation, enabling LLM to provide contextually appropriate suggestions.
6. **Pipeline Recommended by LLM:** The LLM analyzes the prompt and outputs the recommended pipeline, which is stored for further processing.
7. **Parameter Recommendation Prompt:** The system integrates the recommended pipeline with the user's query. While the initial LLM recommender provides suitable pipelines and parameters, a secondary recommender serves as a validation check to ensure the accuracy of parameters and output formats.
8. **Pipeline + Parameters Recommended:** The system finalizes and displays the complete pipeline configuration and explanation, including its structure and parameters, ensuring it meets the user's specific requirements.

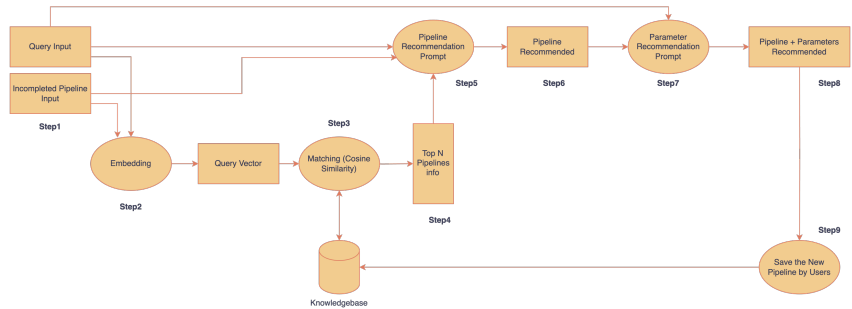

**Fig. 4.** Implementation Workflow of SLEGO LLM Recommendation System

## 4    The Implementation of SLEGO System

This section details the deployment of the SLEGO system, built on Jupyter Notebook [16] and deployable on cloud platforms like AWS [35] or Azure [36]. Python-based components, including a Panel [37] user interface. Data is stored in a shared cloud folder, with SQLite for database management. The recommendation system uses OpenAI's GPT-4o API for enhanced functionality.

### 4.1    Low-Code Analytics UI

The SLEGO user interface is a web application built with Python's Panel library (see **Fig. 5**) and includes the following areas: Data Management Area (Section 1) for managing data files, microservice and knowledge base; Microservice Selection Panel (Section 2) for selecting, assembling, and executing analytics microservices. Parameter Input Panel (Section 3) for customizing microservices through parameter inputs, optionally allows JSON format text input. User Input Area (Section 4) for entering OpenAI API keys and queries for pipeline recommendations; Information Viewer (Section 5) for displaying details on microservices docstring, analytics outputs, and recommendations. On the top, there are tabs to open Microservices Editor (text-based) and Knowledge Base (table-based) Editor for updating, reviewing, and editing microservices and knowledge base.

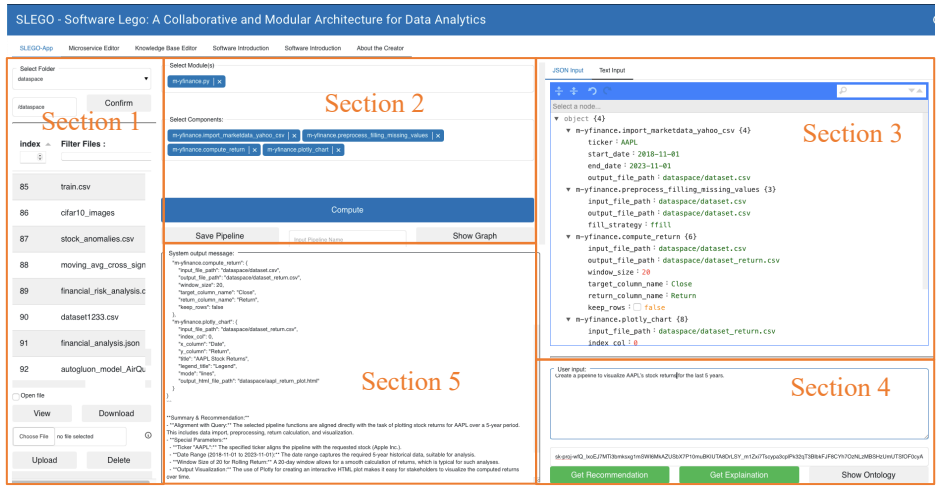

**Fig. 5.** SLEGO Platform User Interface

### 4.2    Microservices Design Guidelines in Practice

In the SLEGO architecture, microservices serve as the core of computational tasks, designed with modularity and ease of integration in mind. Building on prior studies [25], this implementation uses a Python wrapper function approach to construct microservices while adhering to the following guidelines:

- **Python Function Blueprint:** Microservices are encapsulated within Python functions to ensure modularity.
- **Standardized Function Design:** Functions follow a standardized format, featuring input parameters, a descriptive docstring that explains the function's purpose, output details, and a return object (see **Fig. 6**, left box).
- **Flexible Parameters for Analysis:** Each function includes parameters that provide flexibility in the analysis.
- **Data Paths:** Microservices use parameterized input and output file paths, with a centralized dataset folder to streamline data management.
- **Descriptive Docstring:** Each function contains a descriptive docstring to assist both users and the AI recommender in understanding the microservice.
- **Standardized Analytics Pipelines:** Analytics pipeline configuration is formatted in JSON (see **Fig. 6**, right box), handled by the Analytics Pipelines Engine.

```python
def compute_return(input_file_path: str = 'dataspace/dataset.csv',
                   output_file_path: str = 'dataspace/dataset_return.csv',
                   window_size: int = 20,
                   target_column_name: str = 'Close',
                   return_column_name: str = 'Return',
                   keep_rows: bool = False):
    """
    Compute the daily returns of a stock based on the closing price over a given window size.

    Args:
    input_file_path (str): Path to the input CSV file.
    output_file_path (str): Path to save the output CSV file.
    window_size (int): The number of days over which to calculate the percentage change.
    target_column_name (str): The name of the column from which to calculate returns.
    return_column_name (str): The name of the new column for the calculated returns.
    keep_rows (bool): If False, rows containing NaN values as a result of the calculation will

    Returns:
    pd.DataFrame: DataFrame with the returns added as a new column.
    """
    # Read the data from the input file
    data = pd.read_csv(input_file_path)

    # Calculate returns and assign them to the specified new column
    data[return_column_name] = data[target_column_name].pct_change(periods=window_size)

    # Handle NaN values based on keep_rows
    if not keep_rows:
        data = data.dropna(subset=[return_column_name])

    # Save the modified DataFrame to a new CSV file
    data.to_csv(output_file_path, index=False)

    return data
```

```json
{"m-yfinance.import_marketdata_yahoo_csv": {
    "ticker": "msft",
    "start_date": "2023-11-12",
    "end_date": "2024-11-11",
    "output_file_path": "dataspace/dataset.csv"
},
"m-yfinance.preprocess_filling_missing_values": {
    "input_file_path": "dataspace/dataset.csv",
    "output_file_path": "dataspace/dataset.csv",
    "fill_strategy": "ffill"
},
"m-yfinance.compute_return": {
    "input_file_path": "dataspace/dataset.csv",
    "output_file_path": "dataspace/dataset_return.csv",
    "window_size": 20,
    "target_column_name": "Close",
    "return_column_name": "Return",
    "keep_rows": false
},
"m-yfinance.plotly_chart": {
    "input_file_path": "dataspace/dataset_return.csv",
    "index_col": 0,
    "x_column": "Date",
    "y_column": "Return",
    "title": "Data Plot",
    "legend_title": "Legend",
    "mode": "lines",
    "output_html_file_path": "dataspace/dataset_plot.html"
}}
```

**Fig. 6.** SLEGO Python Wrapper Function (Left) and Pipeline Configuration (Right) Examples

## 5    Case Studies: Collaborative Analytics Across Diverse Teams

These case studies explore how analytics tasks are implemented using the SLEGO platform. The evaluation emphasizes its ability to foster collaboration and adaptability across diverse teams. All implementations are available in our GitHub repository.

### 5.1    Background

An organisation employs the SLEGO architecture and platform to streamline analytics workflows among financial analysts, fintech developers, machine learning developers, and domain experts. The platform is designed to facilitate collaboration by providing a modular environment where microservices can be efficiently developed, reused, and documented. In this context, a non-technical financial analyst team aims to perform two distinct data analytics tasks: (1) visualizing stock market data and (2) generating machine learning predictions for stock market returns.

The fintech development team, consisting of technical experts, constructed a pipeline for visualizing stock return data following SLEGO's guidelines (**Fig. 7**). This pipeline, uploaded to the SLEGO platform via the Microservices Editor UI, integrates four microservices: importing stock data from Yahoo Finance, preprocessing the dataset, computing stock returns, and generating an interactive HTML visualisation.

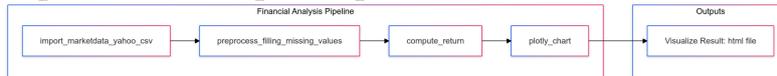

**Fig. 7.** Stock Market Return Visualisation Pipeline

Similarly, the machine learning development team developed an AutoML air quality forecasting pipeline (**Fig. 8**), which they also uploaded to the SLEGO platform. This pipeline comprises five microservices: acquiring data from the UCI API1, preprocessing, splitting the dataset, training an AutoML model, and generating predictions as CSV files alongside performance metrics in a text file.

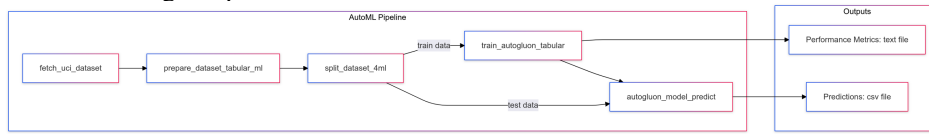

**Fig. 8.** AutoML Air Quality Forecasting Pipeline

Domain experts, such as financial researchers and data science researchers, leverage the SLEGO platform to review these pipelines. They document their functionalities and limitations within the knowledge base via the Knowledge Base Editor UI, enhancing understanding and reusability for users and the recommender system.

### 5.2    Analytics Task 1: Stock Market Data Visualisation

The financial analyst team seeks to utilize SLEGO's GUI to customize and execute pipelines. By selecting microservices through dropdown menu widgets (see **Fig. 5**, Section 2) and modifying parameters in text box widgets (see **Fig. 5**, Section 3), they can adjust visualisations—for example, changing the dataset's date range or ticker symbols. This allows them to tailor the visualisation to specific requirements without needing technical expertise. Finally, the team can access the outcome file via the data management widget (see **Fig. 5**, Section 1).

Alternatively, the analysts can interact with SLEGO's recommender system by submitting a prompt like "Create a pipeline to visualize AAPL's stock returns for the last 5 years." to SLEGO recommender via User Input Area (see **Fig. 5**, Section 4). The recommender generates a pipeline suggestion similar to **Fig 7**, including the necessary parameters and explanation of the selection (see **Fig. 5**, Section 5). This feature enables analysts to quickly generate customized visualisations based on their analytical needs.

---

[1] https://pypi.org/project/uci-dataset/

With the knowledge base and intuitive GUI, the platform allows analysts to build pipeline by selecting microservices and adjusting parameters, empowering them to tailor analyses to their specific needs.

### 5.3    Analytics Task 2: Machine Learning Prediction for Stock Market Data

For more advanced analytics, financial analysts can consult the knowledge base to learn how to manually assemble a predictive pipeline like the one illustrated in **Fig 10**. This pipeline builds upon previously introduced components (**Fig. 7** and **Fig. 8**), enabling analysts to understand the underlying microservices and gradually expand their technical proficiency. If they encounter unfamiliar concepts or prefer more guidance, they can alternatively submit a prompt—such as "AutoML prediction for stock market return"—to SLEGO's LLM-based recommender. This flexibility allows both experienced and less technically inclined users to choose between hands-on learning from the knowledge base or receiving tailored, automated recommendations from the LLM.

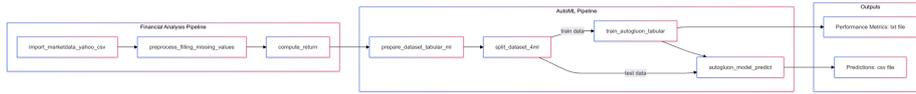

**Fig. 9.** SLEGO Recommender's Suggested Pipeline for AutoML Stock Market Prediction

Analytics Task 2 showcases the system's capability for knowledge discovery by enabling financial analysts to explore advanced machine learning pipelines through the LLM-based recommender. This facilitates the uncovering of new analytical possibilities without necessitating extensive technical skills.

## 6    Discussion

The evaluation of SLEGO through two case studies highlights several key findings that support the effectiveness of its architecture. First, the system demonstrates strong capabilities for component reuse and adaptation, although creating microservices and adding associated metadata to the knowledge base still requires technical expertise. Second, SLEGO enables effective cross-domain tool integration; while configuring these applications demands substantial domain knowledge, the knowledge base helps reduce this gap by providing comprehensive documentation and guidance. Third, the LLM recommendation system successfully lowers technical barriers for non-technical analysts, despite not guaranteeing perfect pipeline recommendations. These use cases demonstrate SLEGO's potential in fostering collaboration among technical teams, domain experts, and non-technical analysts. Technical teams benefit from SLEGO's modular architecture, which enables efficient pipeline development and microservices reuse. Domain experts contribute valuable insights by documenting pipeline functionalities and limitations, enriching the knowledge base to guide future users.

The LLM-based recommender system does not guarantee accurate pipeline recommendations due to several factors, including the quality of knowledge base descriptions, the limitations of the underlying machine learning models, and the inherent challenges in interpreting user queries. In knowledge discovery scenarios like Task 2, where novel or complex pipelines are needed, the recommender is more prone to generating errors compared to its performance with well-established pipelines, as in Task 1. Although the system successfully lowers technical barriers for novice users, it may not always produce fully accurate or functional pipeline configurations, particularly for less familiar or innovative use cases.

These limitations—such as a limited knowledge base of microservices and analytics pipelines, and the absence of quantitative metrics to evaluate recommendation accuracy—highlight opportunities for improvement. Future research could focus on expanding the knowledge base and microservices through crowdsourcing, refining the recommender with real-time feedback loops and multi-agent system, and integrating MLOps processes within the system. Additionally, improving knowledge discovery mechanisms for microservices could enhance the recommender's ability to suggest more accurate and innovative pipelines. By advocating this architecture, we aim to encourage further exploration and development of collaborative analytics systems that bridge the gap between technical and non-technical users.

## 7     Conclusion

By integrating low-code interfaces, reusable microservices, a structured knowledge base, and LLM-based recommendations, SLEGO enables users with diverse expertise levels to construct and execute analytics pipelines. This approach supports component reuse, fosters knowledge sharing, and facilitates collaboration across teams.

The case studies illustrate SLEGO's practical application in supporting a range of tasks, from data visualization to advanced machine learning, highlighting its versatility and effectiveness in different domains. While limitations such as recommender accuracy and knowledge base scalability remain, the results demonstrate SLEGO's potential to address key challenges in accessibility, flexibility, and functionality for general-purpose data analytics.

These findings position SLEGO architecture as a step towards more inclusive and efficient analytics workflows. Future work could focus on expanding the system's knowledge base, improving recommendation accuracy, and exploring new collaborative features to further enhance its utility in diverse organizational contexts. Recognizing its potential to advance research in collaborative and crowdsourced analytics, we have made our implementation openly available as an open-source project on GitHub.

SLEGO: A Collaborative Data Analytics System with Recommender for Diverse Users    15